\documentstyle[aps,version2,epsf,preprint]{revtex} 
\begin{document}
\input{psfig}
\draft
\begin{flushright}
\today \\
FERMILAB-PUB-94/420-E
\end{flushright}
\begin{title}
{\Large {\bf Measurement of the $B_{s}$ Meson Lifetime }}
\end{title}
%
\font\eightit=cmti8
\def\r#1{\ignorespaces $^{#1}$}
\hfilneg
\begin{sloppypar}
\noindent
F.~Abe,\r {13} M.~G.~Albrow,\r 7 D.~Amidei,\r {16} J.~Antos,\r {28}
C.~Anway-Wiese,\r 4 G.~Apollinari,\r {26} H.~Areti,\r 7
P.~Auchincloss,\r {25} F.~Azfar,\r {21} P.~Azzi,\r {20}
% ...
\end{sloppypar}

\vskip .025in
\begin{center}
(CDF Collaboration)
\end{center}

\vskip .025in
\begin{center}
\r 1  {\eightit Argonne National Laboratory, Argonne, Illinois 60439} \\
\r 2  {\eightit Istituto Nazionale di Fisica Nucleare, University of Bologna,
I-40126 Bologna, Italy} \\
\r 3  {\eightit Brandeis University, Waltham, Massachusetts 02254} \\
\r 4  {\eightit University of California at Los Angeles, Los
Angeles, California  90024} \\
\r 5  {\eightit University of Chicago, Chicago, Illinois 60637} \\
\r 6  {\eightit Duke University, Durham, North Carolina  27708} \\
\r 7  {\eightit Fermi National Accelerator Laboratory, Batavia, Illinois
60510} \\
\r 8  {\eightit Laboratori Nazionali di Frascati, Istituto Nazionale di Fisica
               Nucleare, I-00044 Frascati, Italy} \\
\r 9  {\eightit Harvard University, Cambridge, Massachusetts 02138} \\
\r {10} {\eightit University of Illinois, Urbana, Illinois 61801} \\
\r {11} {\eightit Institute of Particle Physics, McGill University, Montreal
H3A 2T8, and University of Toronto,\\ Toronto M5S 1A7, Canada} \\
\r {12} {\eightit The Johns Hopkins University, Baltimore, Maryland 21218} \\
\r {13} {\eightit National Laboratory for High Energy Physics (KEK), Tsukuba,
Ibaraki 305, Japan} \\
\r {14} {\eightit Lawrence Berkeley Laboratory, Berkeley, California 94720} \\
\r {15} {\eightit Massachusetts Institute of Technology, Cambridge,
Massachusetts  02139} \\
\r {16} {\eightit University of Michigan, Ann Arbor, Michigan 48109} \\
\r {17} {\eightit Michigan State University, East Lansing, Michigan  48824} \\
\r {18} {\eightit University of New Mexico, Albuquerque, New Mexico 87131} \\
\r {19} {\eightit Osaka City University, Osaka 588, Japan} \\
\r {20} {\eightit Universita di Padova, Instituto Nazionale di Fisica
          Nucleare, Sezione di Padova, I-35131 Padova, Italy} \\
\r {21} {\eightit University of Pennsylvania, Philadelphia,
        Pennsylvania 19104} \\
\r {22} {\eightit University of Pittsburgh, Pittsburgh, Pennsylvania 15260} \\
\r {23} {\eightit Istituto Nazionale di Fisica Nucleare, University and Scuola
               Normale Superiore of Pisa, I-56100 Pisa, Italy} \\
\r {24} {\eightit Purdue University, West Lafayette, Indiana 47907} \\
\r {25} {\eightit University of Rochester, Rochester, New York 14627} \\
\r {26} {\eightit Rockefeller University, New York, New York 10021} \\
\r {27} {\eightit Rutgers University, Piscataway, New Jersey 08854} \\
\r {28} {\eightit Academia Sinica, Taiwan 11529, Republic of China} \\
\r {29} {\eightit Superconducting Super Collider Laboratory, Dallas,
Texas 75237} \\
\r {30} {\eightit Texas A\&M University, College Station, Texas 77843} \\
\r {31} {\eightit University of Tsukuba, Tsukuba, Ibaraki 305, Japan} \\
\r {32} {\eightit Tufts University, Medford, Massachusetts 02155} \\
\r {33} {\eightit University of Wisconsin, Madison, Wisconsin 53706} \\
\r {34} {\eightit Yale University, New Haven, Connecticut 06511} \\
\end{center}

\begin{abstract}
\noindent
The lifetime of the $B_s$ meson is
measured using the semileptonic decay
$B_s \rightarrow D_s^- \ell^+ \nu X$.
The data sample
consists of 19.3 pb$^{-1}$ of
$p \bar{p}$ collisions at $\sqrt{s} = 1.8$ TeV collected by the
CDF detector at the Fermilab Tevatron collider during 1992-1993.
There are $76 \pm 8$ $\ell^{+} D_s^{-}$ signal
events where the $D_s$ is identified via the decay $D_s^{-}
\rightarrow \phi \pi^{-}$,~$\phi \rightarrow K^+ K^-$.
Using these events, the $B_s$ meson lifetime is determined to be
$\tau_{s} = 1.42^{\ + 0.27}_{\ - 0.23}\ {\rm (stat)} \pm 0.11
\ {\rm (syst)}$ ps.
A measurement of the $B_s$ lifetime in  a low statistics sample of
exclusive $B_s \rightarrow J/\psi \phi$ decays is also presented in this
paper.
\end{abstract}
\pacs{PACS numbers: 13.25.Hw, 14.40.Nd}
%
The lifetime differences between  the  bottom hadrons can probe the $B$-decay
mechanisms which are beyond the simple quark spectator model.
In the case of charm mesons, such differences have been observed to
be quite large ($\tau(D^+)/\tau(D^0)\sim 2.5$).
Among bottom hadrons, the lifetime differences are expected to be
smaller due to the heavier bottom quark mass.
Phenomenological models predict a 5-10\% difference between the $B_u$ and $B_d$
meson
lifetimes and very similar $B_d$ and $B_s$ lifetimes~\cite{theory1}.
This is consistent with the previous measurements of $B_{u,d}$ meson
lifetime~\cite{bu_bd_life}, as well as recent $B_s$ lifetime measurements from
LEP~\cite{lep_bs_life}. It has also been suggested by recent theory
calculations~\cite{theory2} that
the lifetime between the two CP eigenstates produced by mixing of
the $B_s$ and $\overline{B}_s$ may differ by as much as 15\%.
Such an effect may manifest itself as a difference in lifetimes
between the $B_s$ semileptonic decay, which is almost an equal
mixture of the two CP states, and the decay $B_s\rightarrow J/\psi
\phi$, which is expected to be dominated by the CP even state. In this letter,
we first present the measurement of $B_s$ lifetime using the
semileptonic decay ~\cite{foot1}
$B_s \rightarrow D_s^- \ell^+ \nu X$, where the $D_s^-$ is identified via
$D_s^- \rightarrow \phi \pi^- , ~\phi \rightarrow K^+ K^-$. We then describe
briefly a result using the exclusive decay $B_s \rightarrow J/\psi \phi$,
where $J/\psi \rightarrow \mu^{+} \mu^{-}, ~\phi \rightarrow K^{+} K^{-}$.
The data sample for this paper consists of 19.3 $pb^{-1}$ of $p\overline{p}$
collisions at $\sqrt{s}$=1.8 TeV collected by the CDF detector during the
1992-1993 run.
%
%

The CDF detector is described in detail elsewhere~\cite{CDF}.
We describe here only the detector features most relevant to this
analysis.
Two devices inside the 1.4 T solenoid
are used for the tracking of charged particles:
the silicon vertex detector (SVX) and the central tracking chamber (CTC).
The SVX
consists of four layers of silicon microstrip detectors
located at radii between  3.0 and 7.9~cm from
the interaction point
and provides spatial measurements in the $r$-$\varphi$ plane ~\cite{foot2}
with a resolution of 13 $\mu$m, giving a track impact parameter
resolution of about $(13 + 40/p_T)~\mu$m~\cite{svx},
where $p_T$ is the transverse momentum of the track in GeV/$c$.
The transverse profile of the beam is circular and has an RMS
of $\sim 35$ $\mu$m, while the longitudinal beam size is  $\sim 30$ cm.
The CTC is a cylindrical drift chamber containing 84 layers grouped into 8
alternating superlayers of axial and stereo wires.
It covers the pseudorapidity interval $|\eta| < 1.1$,
where $\eta=-\ln[\tan(\theta/2)]$.
The $p_T$ resolution of the CTC combined with the SVX is
$\delta(p_T)/p_T = ((0.0066)^2 + (0.0009 p_T)^2)^{1/2}	$.
Outside the solenoid are electromagnetic (CEM) and hadronic (CHA) calorimeters
($|\eta|<1.1$) that employ a projective tower geometry.
A layer of proportional wire chambers (CES)
is located near shower maximum in the CEM and
provides a measurement of electromagnetic shower profiles
in both the $\varphi$ and $z$ directions.
Two different muon subsystems in the central region are used,
the central muon chambers (CMU) and the central upgrade muon chambers (CMP),
with total coverage  of 80\% for $|\eta| \leq 0.6$. The CMP chambers are
located behind 8 interaction lengths of material.

%
Events containing semileptonic $B_s$ decays were collected using
inclusive electron and muon triggers. The $E_T$ threshold for the principal
single electron trigger was 9~GeV, where $E_T \equiv E\sin(\theta)$ and $E$ is
the electromagnetic energy measured in the calorimeter. The single muon trigger
required a $p_T > 7.5$ GeV/$c$  track in the CTC with matched track segments in
both the CMU and CMP systems.

Offline identification of an electron~\cite{electron} involved measurements
from both the calorimeters and the CTC. Photon conversion electrons were
removed by searching for an oppositely charged track which had a small opening
angle with a primary electron candidate.

A muon candidate was required to be detected by  both the CMU and CMP
chambers to reduce background due to  hadrons that do not interact in the
calorimeter. Good position matching~\cite{cdf_life} was required between track
segments in the muon chambers and an extrapolated CTC track.

%
%
The $D_s^- \rightarrow \phi \pi^-$ reconstruction started with a search for
$\phi$ candidates.
We first defined a search cone around the lepton candidate with a
radius $\Delta R = \sqrt{\Delta\eta^{2} + \Delta\varphi^{2}}$ of 0.8.
Any two oppositely charged tracks
with $p_T >1$~GeV/$c$ within that cone were assigned kaon masses and
combined to form a $\phi$ candidate. No kaon identification was used in the
$\phi$ selection. Each $\phi$ candidate was required to have $p_T(\phi) >
2.0$~GeV/$c$ and a mass within $\pm 8$ MeV/$c^2$ of the world average
$\phi$ mass~\cite{PDGLAT}. The $\phi$ candidate was then
combined with another track of $p_T> 0.8$ GeV/c inside the cone which
had the opposite charge of the lepton (the `right-sign' combination). This
third track was assigned the pion mass. To ensure a good decay vertex
measurement, track quality cuts were imposed on the lepton and
at least two of the three track candidates forming the $D_s$ candidate.
The $K^+$, $K^-$, and $\pi^-$ tracks were then
refit with a common vertex constraint. The confidence level of that fit
was required to be greater than 1\%.
Since the $\phi$ has spin 1 and both the $D_s^-$ and $\pi^-$ are spin 0, the
helicity angle $\Psi$, which is the angle between the $K^+$ and $D_s^-$
directions in the $\phi\ $ rest frame  exhibits a distribution
$dN/d(\cos\Psi) \sim \cos^2\Psi$. A cut $|\cos \Psi| > 0.4\ $ was
therefore applied to suppress the combinatorial background, which
we found to be a flat in $\cos\Psi$ distribution. The mass of the
$\ell D_s$ system was required to be between 3.0 and 5.7 GeV/$c^2$ in order
to be consistent with coming from a $B_s$ decay.
We also applied an
isolation cut $E_T^{\rm iso}/p_T(\phi\pi^-) < 1.2$ on the $D_s^-$
candidate, where $E_T^{\rm iso}$ is a sum of transverse energy within a cone of
radius 0.4 in $\eta$-$\varphi$ space around the lepton candidate, excluding the
lepton energy.
This cut eliminated many of the fake $D_s^-$  combinations from high
track multiplicity jets.  Furthermore, we required that the
apparent $D_s^-$ decay vertex ($V_{D_{s}}$) be positively displaced
from the primary vertex along the direction of the $\ell^+D_s^-$
momentum.
Figure~\ref{fig1}a shows the $\phi\pi^-$ invariant mass distribution
for the `right-sign' lepton-$D_s$ combinations. A $D_s$ signal with mean
of $1.967$ GeV/$c^2$ and width of $5.4$ MeV/$c^2$ is observed. Evidence
of the Cabibbo suppressed $D^- \rightarrow \phi \pi^-$ decay is also
present. No enhancement is seen in the corresponding
distribution for the `wrong-sign' combinations (Figure~\ref{fig1}b).
%
%
We select a signal sample using a $D_s^-$ mass window of 1.953 to 1.981
GeV/$c^2$. A total of 139 events are found with a background fraction
$f_{bg}= 0.45~\pm 0.01$. The number of $\ell^+ D_s^-$ events above
background in the sample is 76$~\pm$ 8.

There are two possible sources of non-strange $B$ meson decays which can
lead to right-sign $\ell^+D_s^-$ combinations.
The first one is a four
body decay $B_{u,d}  \rightarrow D_s^- {\bf K} \ell^+ {\nu}$, where
${\bf K}$ denotes any type of strange meson. Because of the low probability
of producing $s\overline{s}$ pairs and the
limited phase space, this process is suppressed and has not been observed
experimentally.
The recent ARGUS limit (90\% CL) is ${\rm BR}(B_{u,d}
\rightarrow D_s^- {\bf K} \ell^+ \nu )< 1.2\%$~\cite{exp_limit}. Also, a
theoretical analysis based on the `resonance model' yields
${\rm BR}(B_{u,d} \rightarrow D_s^- {\bf K} \ell^+ \nu ) \leq 0.025 \times
{\rm BR}(B_d \rightarrow \ell^+\nu X)$~\cite{limit}. Using the latter
result and our estimated  efficiency, we expect less than 2.6\% of our
$\ell^+D_s^-$ combinations from this source.
The second process is $\overline{B}_{u,d} \rightarrow  D_s^{-} D X,
D \rightarrow \ell^+ {\nu} X$, where $D$ is any charmed meson. This decay
produces softer and less isolated leptons than that from $B_s$ semileptonic
decay and therefore the acceptance for this source relative to the signal is
quite small ($\sim$ 2.6\%). Using the ${\rm BR}(B\rightarrow D_s X)$~\cite
{cleo,argus} and the semileptonic branching ratios of $D^{0}$ and
$D^+$~\cite{PDGLAT}, we estimate the fraction of this type of background
is less than 3\%. In addition, we also considered the background from
$c{\overline c}$ production where a $D_s^- D$ pair is produced. Monte
Carlo predicts the background fraction from this type of source to be $< 7\%$.
In summary, the contribution of all above physics backgrounds is quite small
compared to the combinatorial background. We will consider them as a source
of systematic uncertainty for the $B_s$ lifetime measurement.

%
%
The secondary vertex where the $B_s$ decays to a lepton and a $D_s^-$
(referred to as $V_{B_s}$) is obtained by
intersecting the trajectory of the lepton track with the flight path
of the $D_s^-$ candidate.
The transverse decay length $L$ is defined as the displacement
in the transverse plane of
$V_{B_s}$ from the primary vertex projected onto the direction of the
$p_T(\ell D_s)$. This is our best estimator of the $B_s$ direction.
The effect of the unknown $B_s$ relativistic boost can be partially removed
event-by-event
with the factor $p_T(\ell D_s)/M(B_s)$ (where $M(B_s) = 5.37$~GeV/$c^2$
{}~\cite{thesis}) and leads to a corrected decay length
\begin{eqnarray}
\xi  = {\frac{L \cdot M(B_s)}{p_T(\ell D_s)} },
\end{eqnarray}
which is referred to as the `proper decay length'.
A residual correction between
$p_T(\ell D_s)$ and $p_T(B_s)$
is done statistically
by convoluting a Monte Carlo distribution of the $p_T$ correction factor
K$ = p_T(\ell D_s)/p_T(B_s)$
with an exponential decay distribution in the lifetime fit. The K
distribution has an average value of 0.86 and an RMS of 0.11 and is
approximately constant as a function of $p_T(\ell D_s)$.
To model the  proper decay length distribution of the background events
contained in the signal sample, we define a background sample which
consists of the right-sign events from the $D_s^-$ sidebands (1.885-1.945
and 1.990-2.050 GeV/$c^2$) and the wrong-sign events
from the interval 1.885-2.050 GeV/$c^2$.

The proper decay length distribution (Figure~\ref{fig2}) is fit using an
unbinned maximum
log-likelihood method. Both the $B_s$ lifetime and the background shape
are determined in a simultaneous fit using the signal and background
samples. Thus the likelihood function ${\cal L}$ is a combination of two parts:
\begin{eqnarray}
{\cal L} & = & \prod^{N_S}_i[(1-f_{bg}){\cal F}^i_{Sig}
 + f_{bg}{\cal F}^i_{bg} ]
\cdot \prod^{N_B}_j{\cal F}^j_{bg},
\end{eqnarray}
where $N_S$ and $N_B$ are the number of events in the signal and background
samples.
The signal probability function ${\cal F}_{Sig}$ consists of a normalized
decay
exponential function (defined for only positive decay lengths and
symbolized by ${\cal E}_{+}$) convoluted with the K
distribution  and a Gaussian resolution function ${\cal G}$:
\begin{eqnarray}
{\cal F}^i_{Sig}(c\tau,{\bf s}) = {\cal E}_{+}(-K x,c\tau)\otimes K^{\it dist}
\otimes {\cal G}(\xi^i-x,{\bf s}\sigma^i),
\end{eqnarray}
where $\xi^i$ is the measured proper decay length with uncertainty
$\sigma^i$ (typically $100\mu$m) and
$x$ is the true proper decay length.
The scale factor ${\bf s}$ accounts for the underestimation of the decay
length error.  The background is parameterized by a Gaussian centered at zero,
symmetrical positive and negative exponential tails, and a positive
decay exponential to characterize the heavy flavor decay in the background
sample.

The best fit values of $c\tau$ and $s$ are found to
be $426 ^{~+~80}_{~-~68}~\mu$m and $1.4~\pm 0.1$ respectively.
Figure~\ref{fig2}a shows the proper decay length distribution
of the signal sample with the result of the fit superimposed.
The same distribution of the background samples is shown in
Figure~\ref{fig2}b.
As a consistency check, we also fit the $D_s$ lifetime  from the
proper decay length measured from the tertiary vertex $V_{D_s}$ to the
secondary vertex $V_{B_s}$. The result is $c\tau(D_s)= 135 ^{~+~40}_{~-~
30}~\mu$m, which is consistent with the world average value~\cite{PDGLAT}.

Table~I lists all sources of systematic uncertainty considered in this
analysis.
%
Major contributions come from the source of the background shape,
the non-$B_s$ production, and the resolution function.
To model the contribution to the signal from the
combinatorial backgrounds, we
combined the events from three different sideband regions.
There may be some bias
in choosing the correct mixture. We find a $\pm$ 4\% variation in the lifetime
when using each  sideband region individually.
The dominant source of systematics from  non-$B_s$ production was found to be
${\overline B}_{u,d} \rightarrow D_s^- D X$ decays. This mode was
studied using Monte Carlo simulations and the contribution
to the systematic uncertainty in the lifetime
was found to $\pm$ 4\%.
The effect of
the decay length resolution was studied by varying the scale factor
and using an alternative resolution function consisting of two Gaussians,
giving a 3\% systematic uncertainty.

Quoting the statistical and systematic uncertainties separately, we
measure the $B_s$ lifetime using semileptonic decays to be
\begin{eqnarray*}
\tau_{B_s} =
1.42^{~ + 0.27}_{~- 0.23}{\rm~(stat)~} ^{~+ 0.11}_{~- 0.11}
{\rm~(syst)~} {\rm ps}.
\end{eqnarray*}
This result is consistent with the previous world average of $1.34^{~+
0.32}_{~- 0.27}$ ${\rm ps}$ ~\cite{PDGLAT}.

For the exclusive mode measurement, we use the decay chain
$B_{s} \rightarrow J/\psi \phi$,
$J/\psi \rightarrow \mu^{+} \mu^{-}$, $\phi \rightarrow K^{+} K^{-}$.
The data sample and reconstruction techniques used for this
decay channel are similar to those described
in detail elsewhere~\cite{west}. Briefly,
the invariant mass of two oppositely charged muon candidates is
calculated after the tracks are constrained to originate from a common
vertex. $J/\psi$ candidates are selected by requiring the
difference between the dimuon mass and the world average $J/\psi$
mass~\cite{PDGLAT}
to be $< 3 \sigma$, where $\sigma$ is the mass
uncertainty calculated for each dimuon candidate.
The $\phi$ meson selections are the same as reference~\cite{west} but with
$p_T(\phi)> 3.0$ ( rather than 2 GeV/c) to further reduce the background.
To reconstruct $B_s$ meson candidates,
the 4 daughter
tracks are constrained to originate from a common vertex and the
dimuon mass is simultaneously constrained to the world average $J/\psi$ mass.
We require the $\chi^2$ probability for this combined
fit to be $>$ 2\%.
In addition, at least one of the $\mu$ candidates and at least one of the other
tracks must be well measured in the SVX.

Using the measured $p_T(B_s)$, the proper decay length is calculated and the
lifetime of the $B_{s}$ meson is determined by performing a simultaneous
unbinned log-likelihood fit to the entire mass and proper decay length
spectra. The mass distribution is fit to a Gaussian and a flat background.
We model the proper decay length distribution of the background with a
Gaussian centered at zero and positive
and negative exponential functions. The signal is
described by an exponential decay function convoluted with a Gaussian. This fit
determines the mass and lifetime of the $B_{s}$, the signal fraction, and
background shape simultaneously. The proper decay length distribution
is shown in Figure~\ref{fig3}, where we have displayed events within
$\pm$21 MeV/$c^2$ of the $B_s$ mass peak.
The fit
returns $7.9 ^{~+3.6}_{~-1.6}$ signal events and a $B_{s}$ lifetime of
$c\tau_{B_{s}}=520 ^{~+330}_{~-210}~\mu$m.

We estimate a total systematic error of $\pm 20 \mu$m, the dominant
contributions arising from the uncertainty in the parametrization of the
background shape and our understanding of the resolution function.
The $B_s$ lifetime using fully reconstructed $B_s \rightarrow
J/\psi \phi$
decays is measured to be:
\begin{eqnarray*}
\tau_{B_s} = 1.74 ^{~+1.08}_{~-0.69}~({\rm stat.}) ~\pm 0.07~({\rm syst.})
{}~{\rm ps.}
\end{eqnarray*}

In conclusion,
the $B_s$ lifetime has been measured in both the semileptonic
and exclusive decay channels.
At present, the two measurements are consistent with each other
within their quoted uncertainties and are consistent with the results
of the $B_u$ and $B_d$ lifetimes previously measured by CDF.

We anticipate a more significant result in both modes after the ongoing
collider run.

%
%
     We thank the Fermilab staff and the technical staffs of the
participating institutions for their vital contributions. This work was
supported by the U.S. Department of Energy and National Science
Foundation; the Italian Istituto Nazionale di Fisica Nucleare; the
Ministry of Education, Science and Culture of Japan; the Natural
Sciences and Engineering Research Council of Canada; the National
Science Council of the Republic of China; the A. P. Sloan Foundation;
and the Alexander von Humboldt-Stiftung.

\begin{table}[h]
\caption{Semileptonic mode systematic uncertainties.}
\label{tab1}
\begin{tabular}{|c|r|}          \hline
 Systematic Source & Uncertainty \\ \hline
 Background shape    &   4\% \\
 Non-$B_s$ source    &   4\% \\
 Resolution function &   3\% \\
 Boost correction    &   2\% \\
 Decay length cut    &   2\% \\
 Fitting method      &   1\% \\
 Misalignment        &   2\% \\\hline
 Total   &  7\% \\ \hline
\end{tabular}
\end{table}
%
%

\figure{\centerline{\epsfxsize=5.5in\epsffile{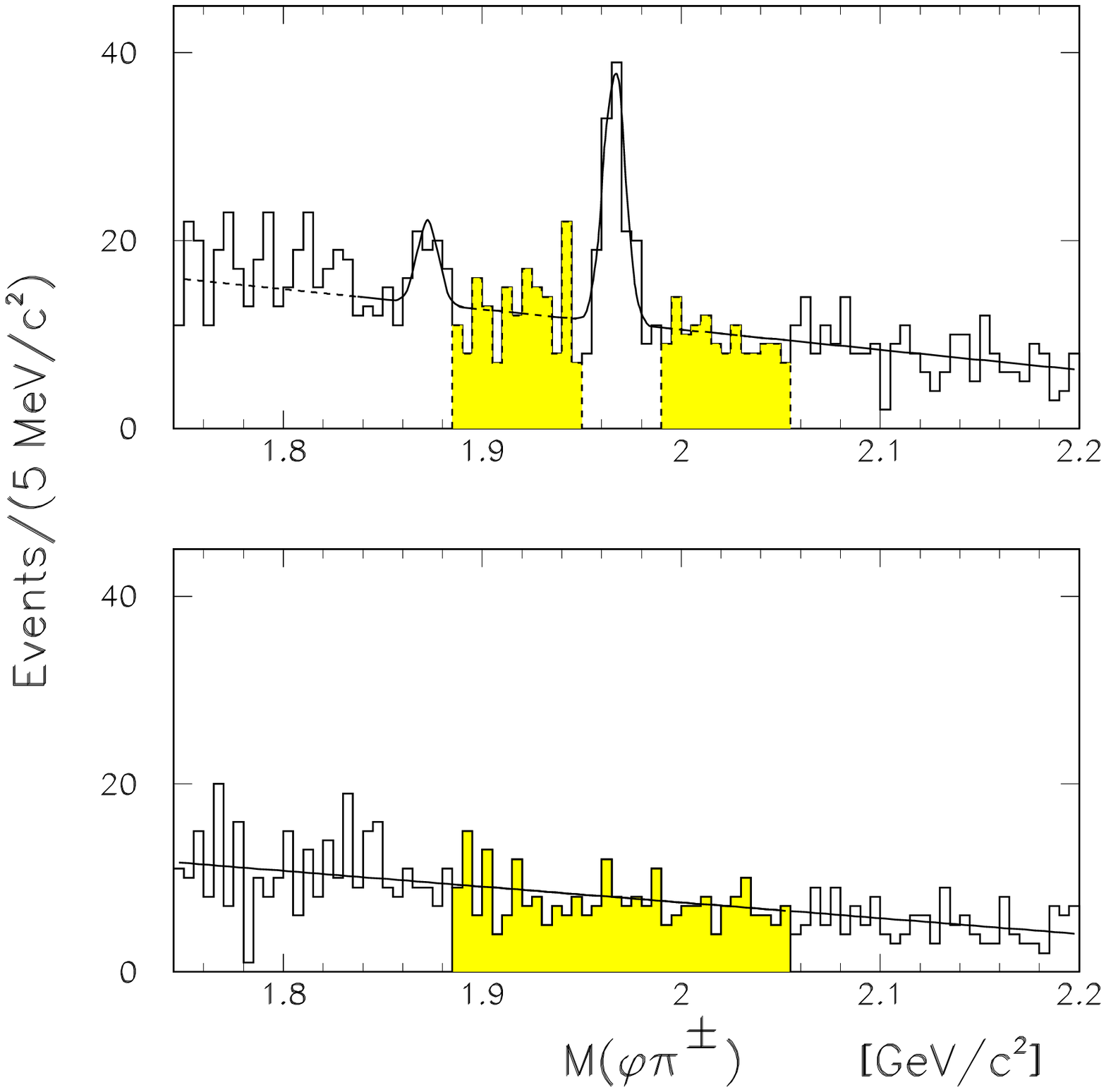}}
The mass distribution of $\phi\pi^{-}$ for (a)
`right-sign' combination $(\phi\pi^{-}\ell^+)$;
(b) `wrong-sign' combination $(\phi\pi^{-}\ell^-)$.The shaded regions
are used for the background sample.
\label{fig1}}
%
%
\figure{\centerline{\epsfxsize=5.5in\epsffile{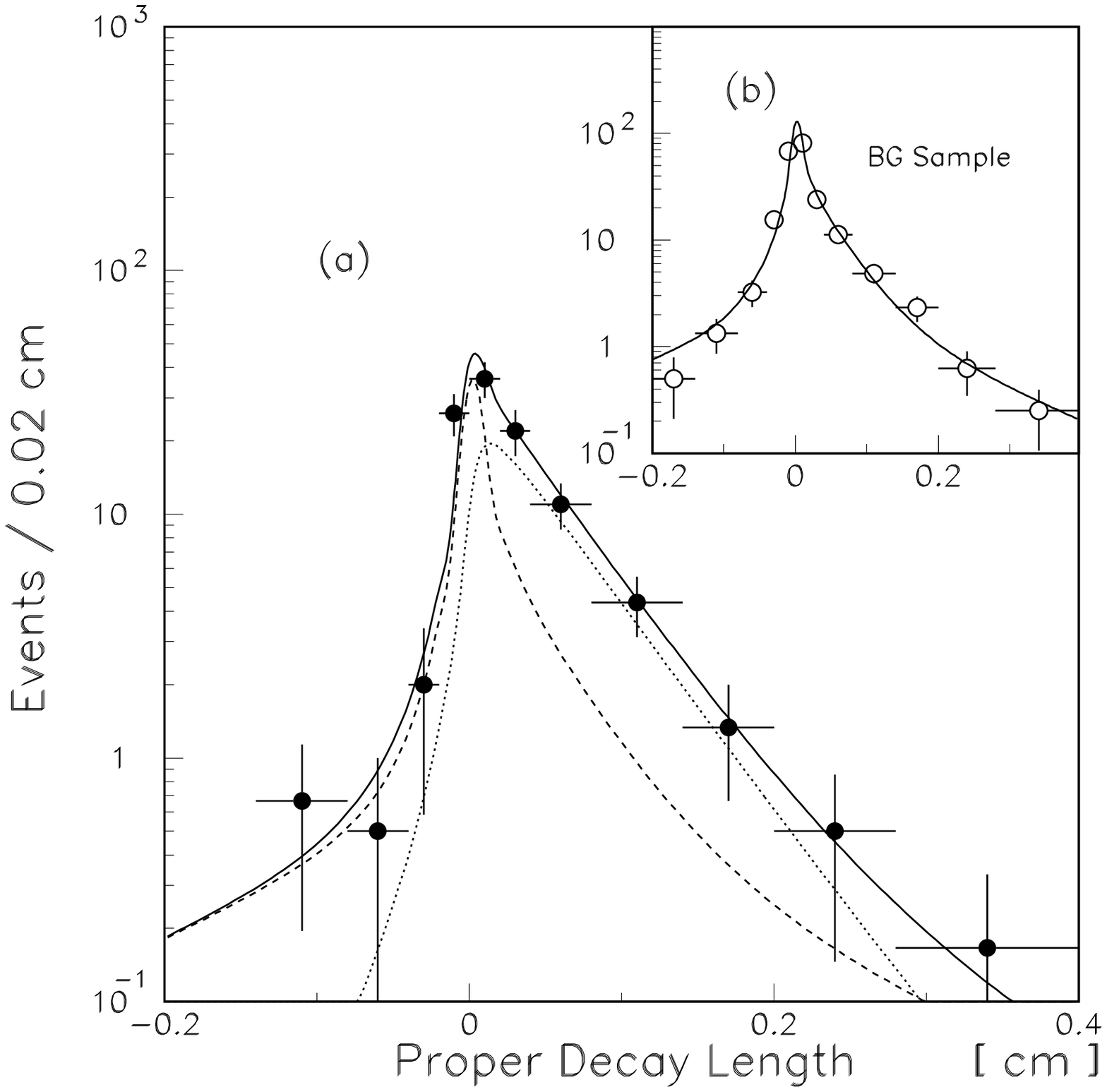}}
(a) Proper decay length distribution for the $\ell^+D_s^-$ signal
sample with a curve (solid) from unbinned log-likelihood fitting of
signal and background. The dashed line represents the contribution
from the combinatorial background. The dotted one represents the signal
contribution. (b) The proper decay length distribution
for the background sample with a curve representing the background
lifetime contribution.
\label{fig2}}
\figure{\centerline{\epsfxsize=5.0in\epsffile{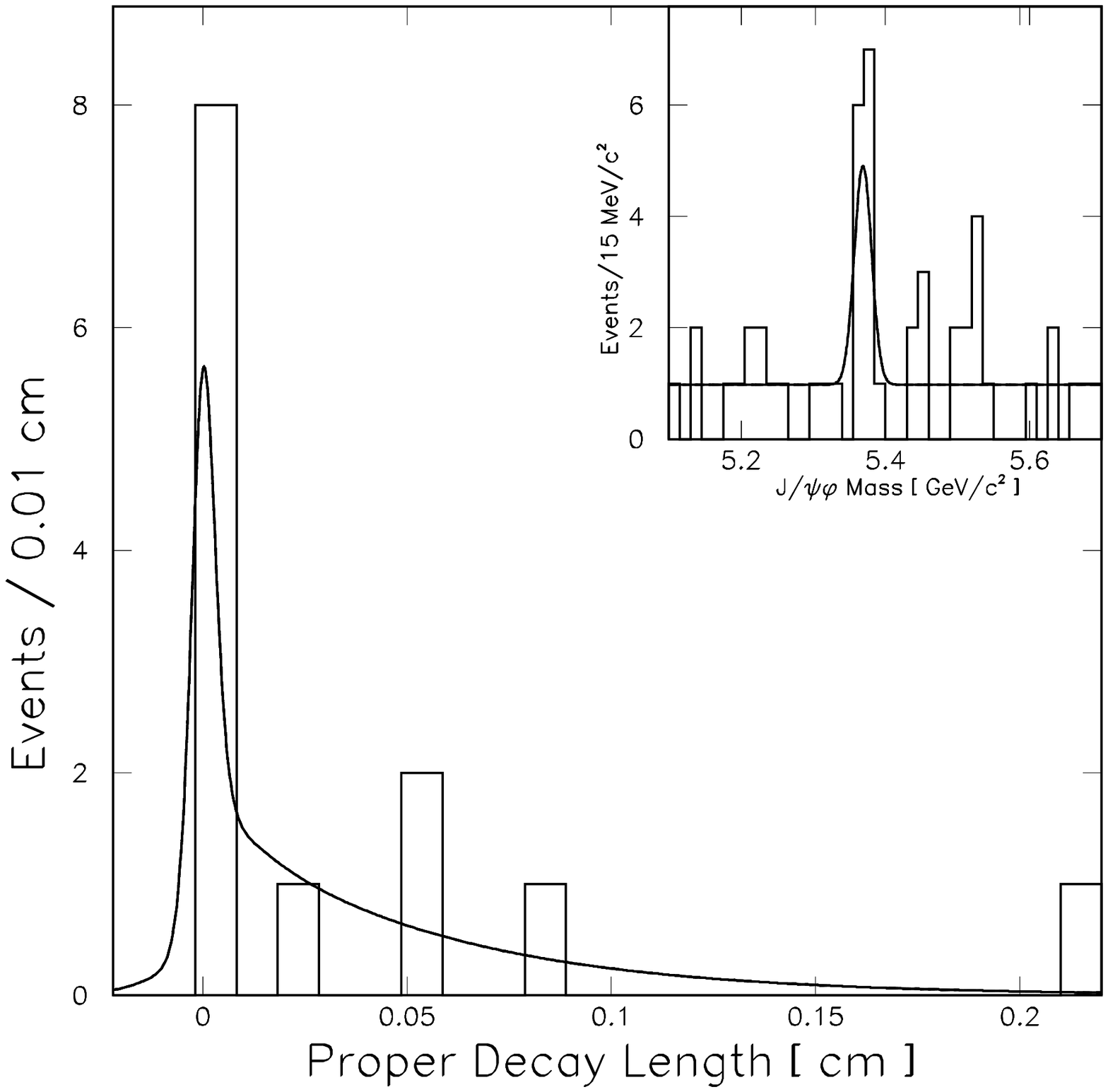}}
$B_s$ lifetime measurement using the $J/\psi \phi$ exclusive mode.
Inset figure is the mass distribution.  The solid curves show the fit results.
\label{fig3}}


\begin{references}
\bibitem{foot2} In CDF, $\varphi$ is the azimuthal angle,
$\theta$ is the polar angle measured from the proton direction,
and $r$ is the radius from the beam axis ($z$-axis).
\end{references}
\end{document}